\documentclass[titlepage,aps,showpacs,prd,twocolumn,groupedaddress,superscriptaddress,floatfix]{revtex4-2}
\usepackage[utf8]{inputenc}
\usepackage{graphicx}
\usepackage{dcolumn}
\usepackage{bm}
\usepackage{amsmath}
\usepackage{amsfonts}
\usepackage{comment}
\usepackage{upgreek}
\usepackage{tabularx}
\usepackage[table]{xcolor}
\usepackage{soul}
\usepackage{marginnote}
\usepackage{amssymb}
\usepackage{multirow}
\usepackage{graphicx}

\setstcolor{red}

\usepackage[hidelinks]{hyperref}

\begin{document}

\title{Invariance property in inhomogeneous scattering media with refractive-index mismatch}

\author{Federico Tommasi} 
\email{federico.tommasi@unifi.it}
\author{Lorenzo Fini} 
\author{Fabrizio Martelli} 
\author{Stefano Cavalieri} 
\affiliation{Dipartimento di Fisica e Astronomia, Universit$\grave{a}$
di Firenze, Via Giovanni Sansone 1 I-50019 Sesto Fiorentino, Italy.} 
\begin{abstract}
The mean path length invariance property is a very important property of scattering media illuminated by an isotropic and homogeneous radiation. Here we investigate the case of inhomogeneous media with refractive index mismatch between the external environment and also among their subdomains. The invariance property remains valid by the introduction of a correction, dependent on the refractive index, of the mean path length value. It is a consequence of the stationary solution of the radiative transfer equation in a medium subjected to an isotropic and homogeneous radiance. The theoretical results are in agreement with the reported results for numerical simulations for both the three-dimensional and the two-dimensional media.

\end{abstract}

\maketitle

\section{Introduction}
Light scattering in optical media is a research field that involves a wide range of different theoretical and experimental issues
, such as  light propagation in biological tissue  \cite{Jacques_2013,Durduran_2010,Leff2008,Zaccanti:03}, random lasers \cite{rl2,rl3,nostro1,PhysRevA.91.033820,PhysRevA.98.053816}, Anderson localization \cite{and_loc,and1}, anomalous diffusion \cite{lf1,LF2,PhysRevA.99.063836} and replica symmetry breaking phenomena \cite{rsb1,rsb3,rsb4}. Moreover, concerning the applications, light diffusion has also been studied in optical sensing \cite{doi:10.1063/1.1782259,0031-9155-56-2-N01,sensore,sens_scat,Tommasi:18} and to find new design for solar cells, in order to enhance the absorption effects \cite{atwater,green,Mupparapu:15,Pratesi:13,Bigourdan:19}; in this regard, it has been reported that a small amount of scattering can promote the absorption in a thin slab subjected to mono-directional illumination \cite{TOMMASI2020124786}.

A fascinating and also, at first glance,  counterintuitive invariance property  holds for the mean total path length $\langle L \rangle$ spent by light propagating inside a disordered medium, and in general also by any kind of particle under diffusion and random walk. Under the hypothesis of isotropic and uniform illumination upon the surface of the medium, $\langle L \rangle$ only depends on the geometry of the medium, whatever is its scattering strength, usually described by the scattering coefficient $\mu_s$. Such an invariance property (IP) \cite{blanco,dubi,Pierrat17765}, also known as Cauchy formula, is a generalization, in the case of scattering, of the mean chord theorem, also described by Dirac in the context of nuclear physics \cite{Dirac}. Recently, the experimental observation of the IP has been reported in the optical case  \cite{Savo765} and in the context of biology by studying the random walks of bacteria in a complex structure \cite{batteri}. 

The IP  remains valid also considering bounded domains of different dimensions; in the three-dimensional and in the two-dimensional case, $\langle L \rangle$  is proportional to the volume-surface ratio, and the surface-perimeter ratio  respectively,
\begin{equation}
\langle L \rangle_{3D}=4\frac{\mathcal{V}}{\mathcal{S}},
\label{L_ip_3D}
\end{equation}
\begin{equation}
\langle L \rangle_{2D}=\pi\frac{\mathcal{S}}{\mathcal{P}}.
\label{L_ip_2D}
\end{equation}
Geometrical details  and  strength of the disorder do not play any role in determining the value of $\langle L \rangle$. Moreover, generalizations of the IP property have been reported in the case of inhomogeneous random media \cite{Mazzolo_2004}, stochastic mixture media \cite{Zoia_2019} and  in  the presence  of absorption and branching  \cite{Zoia_2012,de_Mulatier_2014}.

Here we consider the usual case where the scattering process can be described by the radiative transfer equation (RTE).


The IP remains valid also in case of a infinitely extending medium, such as a thin slab, and with different values of the asymmetry factor $g$ of the scattering function \cite{TOMMASI2020124786}.

In this paper, we investigate how such  property - with suitable modifications - still holds in the case of refractive index mismatch between the disordered medium and the external environment. We have investigated the case of inhomogeneous media with refractive index mismatch between the external environment and also among their subdomains,  both in 3D and in 2D geometry.
\begin{figure}
\centerline{{\includegraphics[width=1.0\columnwidth]{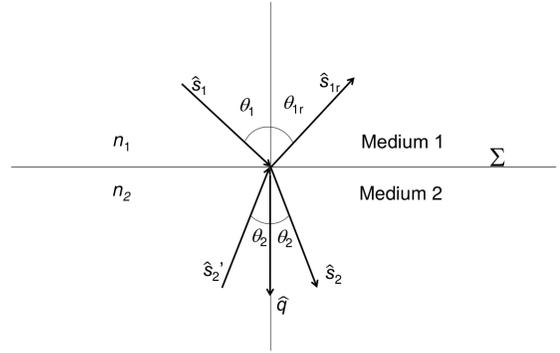}}}
\caption{Schematic of a plane boundary between two media (2D section) and used symbols: $n_1$ and $n_2$ are the refractive indices of medium 1 and medium 2, and $q$ is the normal direction to the boundary and  $\Sigma$ the interface boundary.}
\label{FIG_Plane_Boundary1}
\end{figure}

\section{Theory}
In this paper, we have studied media where scattering plays the leading role and then where interference effects due to refractive index mismatches are destroyed or strongly reduced. Our analytical and numerical results are, indeed, valid in such condition.
Interference effects  due to the bulk structure should be considered near the condition of Anderson localization, i.e.~ $\mu_s\,\lambda=\lambda/l_s \approx 1$, where $l_s$ is the scattering mean free path. However, our simulation, and analytical results, are indeed conceived for the optical or near infrared spectral region where $l_s/\lambda\gg 1$.

To express the boundary condition between two scattering media 1 and 2, one has to take into account that  light intensity can be transferred reciprocally between the media (see Fig.~\ref{FIG_Plane_Boundary1}). The physical condition is that the power flowing in the generic direction $\hat{s}_{2}$ per unit of time and surface at $\vec{r}$ on the  boundary $\Sigma $  must be equal to the sum of the fraction of the transmitted power per unit of time and surface (from medium 1 to medium 2) around the direction $\hat{s}_{1}$  and the fraction of the reflected  power per unit of time and surface around the out-coming direction $\hat{s}_{2}'$ where the directions $\hat{s}_{2}$ and $\hat{s}_{1}$ are related by Snell's law, whereas  $\hat{s}_{2}'$ is the mirror image direction of $\hat{s}_{2}$ (with respect to $\Sigma $).

The relation can be written in full generality in terms of radiance (power per unit of time, surface and solid angle) $I_{2}(\vec{r},\hat{s}_{2},t)$, $I_{2}(\vec{r},\hat{s}_{2}',t)$ and $I_{1}(\vec{r},\hat{s}_{1},t)$ for any point $\vec{r}$ belonging to $\Sigma$ and for  any incoming  direction $s_2$  from medium 1 to medium 2 as
\begin{equation}
\begin{split}
&I_{2}(\vec{r},\hat{s}_{2},t)\cos\theta_2d\Omega_2=\\ 
&=\left[1-R_{F12}(\theta _{1})\right]I_{1}(\vec{r},\hat{s}_{1},t)\cos\theta_{1}d\Omega_1 \\&+R_{F21}(\theta _{2})I_{2}(\vec{r},\hat{s}_{2}',t)\cos\theta_2d\Omega_2~, 
\end{split}
 \label{BC RTE PlaneBound}
\end{equation}

with $\theta_2=\arcsin(\frac{n_1}{n_2}\sin\theta_1)$, and $R_{F12}$ reflection coefficient for the transfer from medium 1 to medium 2 for non-polarized radiation, $R_{F21}$ reflection coefficient for the transfer from medium 2 to medium 1 and with $d\Omega_1$ and $d\Omega_2$ infinitesimal solid angles around the directions $\hat{s}_{1}$ and $\hat{s}_{2}$, respectively. Then, taking into account the Snell's law (and its differential) one obtains:
\begin{equation}
\begin{split}
&I_{2}(\vec{r},\hat{s}_{2},t)-R_{F21}(\theta _{2})I_{2}(\vec{r},\hat{s}_{2}',t)=\\ &\left[
1-R_{F12}(\theta _{1})\right]I_{1}(\vec{r},\hat{s}_{1},t)\left(\frac{n_2}{n_1}\right)^2. \end{split} \label{BC RTE PlaneBound_3}
\end{equation}
It is worth noting that $R_{F21}(\theta _{2})=R_{F12}(\theta _{1})$ for all the incoming directions for which a refracted beam exists \cite{jw}. 

The procedure can be applied also to a 2D geometry where the radiance is defined as the power per unit of time, length, and angle, resulting in
\begin{equation}
\begin{split}
&I_{2}(\vec{r},\hat{s}_{2},t)\cos\theta_2d\theta_2=\\&=\left[   
1-R_{F12}(\theta _{1})\right]I_{1}(\vec{r},\hat{s}_{1},t)\cos\theta_{1}d\theta_1+\\&+R_{F21}(\theta _{2})I_{2}(\vec{r},\hat{s}_{2}',t)\cos\theta_2d\theta_2~, \end{split} \label{BC RTE PlaneBound_4}
\end{equation}

Proceeding similarly to the previous case, one obtains the 2D result,

\begin{equation}
\begin{split}
&I_{2}(\vec{r},\hat{s}_{2},t)-R_{F21}(\theta _{2})I_{2}(\vec{r},\hat{s}_{2}',t)=\\&=\left[
1-R_{F12}(\theta _{1})\right]I_{1}(\vec{r},\hat{s}_{1},t)\left(\frac{n_2}{n_1}\right).  \label{BC RTE PlaneBound_5}
\end{split}
\end{equation}

\begin{figure}
\centerline{{\includegraphics[width=1.0\columnwidth]{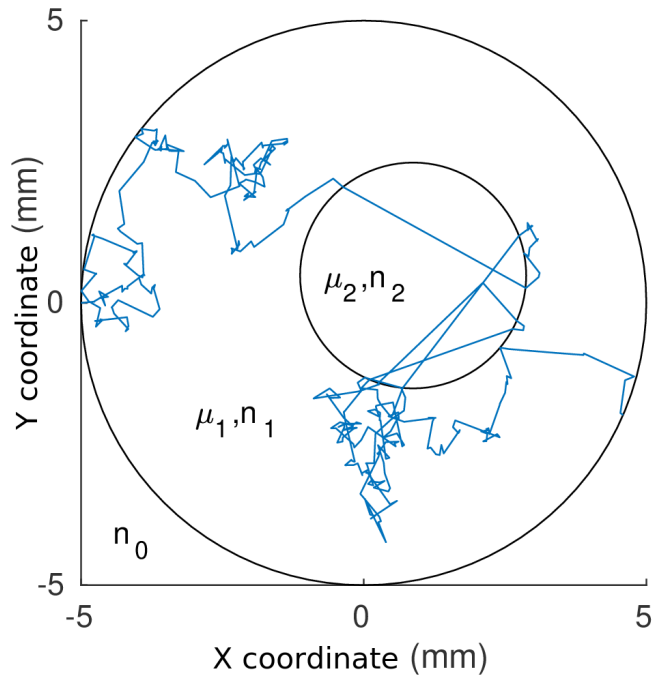}}}
\caption{(color online) Example of trajectory in the MC simulation of a 2D random walk inside a bounded circular domain with a circular decentered inhomogeneity. The starting position is located in the boundary of the larger medium, at coordinates (-5,0) mm. The refractive index $n_0$ of the external environment is 1. The largest medium has refractive index $n_1=2$ and scattering coefficient $\mu_1=5$ mm$^{-1}$, whereas in the inhomogeneity $n_2=1$ and $\mu_2=0.1$ mm$^{-1}$.   }
\label{es_rw_2D}
\end{figure}

\subsection{Boundary of a non-absorbing medium subjected to Lambertian illumination}
The case of the continuous wave Lambertian illumination is a significant case where the boundary conditions assume a particularly simple form and it is the main hypothesis of the invariance property in presence of scattering. Let it be a uniform isotropic illumination applied at the external boundary $\Sigma$ of a finite, uniform, scattering and non-absorbing medium of volume $V$:  $I_{1}(\vec{r},\hat{s}_{1},t)=I_1$ for $\forall \vec r\in\Sigma$. For such a condition, it can be verified that at the interior of the volume a homogeneous and isotropic radiance of the form

\begin{equation}
I_{2}(\vec{r},\hat{s}_{2})=I_{2}=I_{1}\left(\frac{n_2}{n_1}\right)^2  \label{BC RTE PlaneBound_6}
\end{equation}

is a stationary solution of the RTE equation and satisfies the boundary conditions of  Eq.~(\ref{BC RTE PlaneBound_3})    with $R_{F21}(\theta _{2})=R_{F12}(\theta _{1})$.

The analog result of Eq.~(\ref{BC RTE PlaneBound_6})  is obtained in 2D

\begin{equation}
I_{2}(\vec{r},\hat{s}_{2})=I_{2}=I_{1}\left(\frac{n_2}{n_1}\right).  \label{BC RTE PlaneBound_7}
\end{equation}

For a more general case, let us now assume that the volume $V$ is inhomogeneous in scattering with several discrete sub-volumes $V_i$. Once the incoming radiance $I_1$  is uniformly and isotropically incident on the surface $\Sigma$ containing the total volume (hypothesis of the Invariance Property), the solution of the RTE in any volume is again a uniform radiance $I_i$ in any sub-volumes $V_i$. For such a solution the connection of the radiances between adjacent volumes $V_i$ and $V_j$ will be then equal to that of Eqs.~\ref{BC RTE PlaneBound_6} and \ref{BC RTE PlaneBound_7}, that is as follows:

\begin{equation}
I_{j}=I_{i}\left(\frac{n_j}{n_{i}}\right)^2,  \label{BC RTE PlaneBound_8}
\end{equation}
in 3D and 
\begin{equation}
I_{j}=I_{i}\left(\frac{n_j}{n_{i}}\right),  \label{BC RTE PlaneBound_9}
\end{equation}
in 2D.

\begin{figure}
\centerline{{\includegraphics[width=1.0\columnwidth]{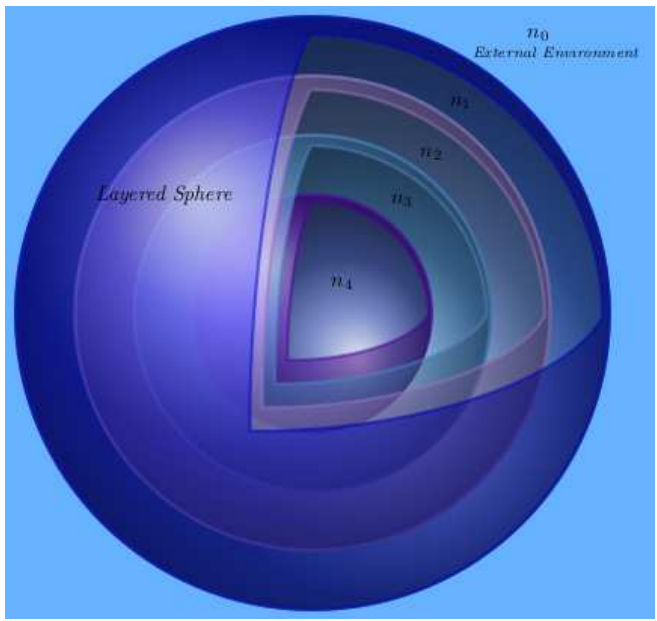}}}
\caption{(color online) Structure of the layered sphere, with the refractive index and the scattering coefficient of each layer. The radiation impinges, from the external environment to the surface of the outermost layer, in  the Lambertian way.}
\label{lsphereA}
\end{figure}

With reference to Figs.~\ref{es_rw_2D} and \ref{lsphereA} it is interesting to note that the radiance in any sub-volume can be expressed by means of the radiance $I_1$ incoming on the more external surface $\Sigma$ , the external index of refraction  $n_1$ and the refraction index $n_i$ of the specific sub volume:

\begin{equation}
I_{i}=I_{1}\left(\frac{n_i}{n_{1}}\right)^2,  \label{BC RTE PlaneBound_10}
\end{equation}

in 3D and

\begin{equation}
I_{i}=I_{1}\left(\frac{n_i}{n_{1}}\right),  \label{BC RTE PlaneBound_11}
\end{equation}
in 2D.


\subsection{Invariance property: inhomogeneous case for the refractive index}

Eqations (\ref{L_ip_3D}) and (\ref{L_ip_2D}) can be  generalized to the case of media with inhomogeneous  refractive index. Let us assume to have an inhomogeneous medium such as  described in the previous section.  Following the approach of Blanco and Fournier \cite{blanco} we assume to have a very small absorption coefficient $\mu_a$ throughout the volume and we equate the power  absorbed in the volume $V$ to the absorption of the power impinging the external surface from outside (let us label with ``\textit{e}'' the quantity referred to the outside).

For the  last quantity, assuming $\mu_a$ is constant in all the volumes, in the limit $\mu_a \rightarrow 0$, we have as follows:
\begin{equation}
P_A=   \pi S I_e \int\limits_0^{+\infty} [1-\exp(-\mu_a L)]P(L)dL   = \pi S I_e \mu_a \langle L \rangle.  \label{PA1}
\end{equation}
where $P(L)$ is the probability density function to have a path-length $L$ within the whole volume $V$ when a Lambertian illumination is impinging on the external boundary $\Sigma$ of area $S$.

As stated above the power absorbed is also given by the  absorption in the volume:
\begin{equation}
P_A=  \mu_a  \int\limits_{V}d\vec{r} \int\limits_{4\pi} I(\vec{r},\hat{s})d\hat{s}=\mu_a\sum\limits_{i=1}^N\int\limits_{V_i}d\vec{r} \int\limits_{4\pi} I_i(\vec{r},\hat{s})d\hat{s}.  \label{pALBInvarianceInhom}
\end{equation}

Using the expression of the radiance obtained in the previous section we have as follows:

\begin{equation}
P_A=\mu_a\sum\limits_{i=1}^N\int\limits_{V_i}d\vec{r}\int\limits_{4\pi} I_i(\vec{r},\hat{s})d\hat{s} . =4\pi I_e \sum\limits_{i=1}^N V_i\left(\frac{n_i}{n_e}\right)^2 \label{pALBInvarianceInhom2}
\end{equation}

and then:

\begin{equation}
 \langle L \rangle=4    \frac{\sum\limits_{i=1}^N V_i\left(\frac{n_i}{n_e}\right)^2}{S} .  \label{pALBInvarianceInhom4}
\end{equation}
Such an equation has been heuristically used in the paper of Savo \emph{et al.} \cite{Savo765} limited to the case of homogeneous medium ($N=1$).

In the case of 2D geometry, we follow a similar approach. We equal the power absorbed in the surface $S$ to the absorption of the power impinging the perimeter $P$ from outside.
The latter in the limit $\mu_a \rightarrow 0$ is given by:

\begin{equation}
P_A=  2 P I_e \int\limits_0^{+\infty} [1-\exp(-\mu_a L)]p_L(L)dL   = 2 P I_e \mu_a \langle L \rangle.  \label{PAsurface1}
\end{equation}

whereas the power absorbed in the surface S is as follows:

\begin{equation}
P_A=  \mu_a \int\limits_{S} d\vec{r} \int\limits_{2\pi} I(\vec{r},\hat{s})d\hat{s}=\mu_a\sum\limits_{i=1}^N\int\limits_{S_i}d\vec{r}\int\limits_{2\pi} I_i(\vec{r},\hat{s})d\hat{s}.  \label{pALBInvarianceInhomsurface}
\end{equation}

Using the expression of the radiance in 2D obtained in the previous section, we have as follows:

\begin{equation}
P_A=\mu_a\sum\limits_{i=1}^N\int\limits_{S_i}d\vec{r}\int\limits_{2\pi} I_i(\vec{r},\hat{s})d\hat{s} . =2\pi I_e \sum\limits_{i=1}^N S_i\left(\frac{n_i}{n_e}\right) \label{pALBInvarianceInhom2surface}
\end{equation}
and then:

\begin{equation}
 \langle L \rangle=\pi    \frac{\sum\limits_{i=1}^N S_i\left(\frac{n_i}{n_e}\right)}{P} .  \label{pALBInvarianceInhom4surface}
\end{equation}

Thus, the invariance property remains valid also in inhomogeneous media with variations of refractive index. The above demonstration also holds  in the presence of ``cavities''. In the following, we show the results of numerical computations in some cases and we compare them with the analytical formula of Eqs.~(\ref{pALBInvarianceInhom4}) and (\ref{pALBInvarianceInhom4surface}).

\section{Numerical Simulations}\label{sim_toeria}

We simulated light diffusion by generating a large number of random walkers and letting them  propagate in the sample volume. We do not take into account interference effects because, due to the stochastic character of the diffusion process, phase information can be neglected.

The numerical simulations in the 3D case are based on a well tested Monte Carlo (MC) code, developed during the 1990s \cite{Zaccanti:91,Contini:97,Sassaroli:98}. The robustness of the core of the code, i.e.\ the generation of trajectories, is checked to be in excellent agreement with exact analytical expressions \cite{Zaccantietal94}. Each step of length $\ell$ of a trajectory is randomly drawn by the exponential probability density function:
\begin{equation}
p\left(\ell\right)=\mu_s\exp{\left[-\mu_s\ell\right]}.
\label{LB}
\end{equation}
Given a uniformly distributed random number $\xi\in[0,1]$, each step of the random walk is obtained by the usual inversion of the cumulative distribution associated to the Eq.~(\ref{LB}),
\begin{equation}
\ell\left(\xi\right)=-\mu_s^{-1}\,\ln{\left[1-\xi\right]}
\label{LB2}
\end{equation}
The used scattering function is the Henyey-Greenstein \cite{hg2}, here considered with $g=0$.

\begin{table}
\begin{tabular}{cc|c|c|c|c|c|c|c|}
\cline{3-7}
& & \multicolumn{5}{ c| }{\bf{Layer index $i$}} \\ \cline{3-7}
& & \bf{0} & \bf{1} & \bf{2} & \bf{3} & \bf{4}  \\ 
\cline{3-7}\hline

\multicolumn{1}{ |c  }{\multirow{1}{*}{\bf{A}} } &
\multicolumn{1}{ |c| }{$n_i$} & 1.00 & 2.00 & 1.00 & 2.00 & 1.00     \\ \cline{2-7}
\hline
\multicolumn{1}{ |c  }{\multirow{1}{*}{\bf{B}} } &
\multicolumn{1}{ |c| }{$n_i$} & 2.00 & 1.00 & 2.00 & 1.00 & 2.00   \\ \cline{2-7}
\hline
\multicolumn{1}{ |c  }{\multirow{1}{*}{\bf{C}} } &
\multicolumn{1}{ |c| }{$n_i$} & 1.00 & 1.25 & 1.50 & 1.75 & 2.00   \\ \cline{2-7}
\hline
& & \multicolumn{5}{ c| }{\bf{Layer index $i$}} \\ \cline{3-7}
&  & \bf{0} & \multicolumn{4}{ c| }{\bf{1,2,$\dots$,100}}  \\ 
\cline{3-7}\hline
\multicolumn{1}{ |c  }{\multirow{1}{*}{\bf{D}} } &
\multicolumn{1}{ |c| }{$n_i$} & 1.00 & \multicolumn{4}{ c| }{1.01,1.02,$\dots$,2.00}  \\ 
 \cline{1-7}\hline
\multicolumn{1}{ |c  }{\multirow{1}{*}{\bf{E}} } &
\multicolumn{1}{ |c| }{$n_i$} & 2.00 & \multicolumn{4}{ c| }{1.99,1.98,$\dots$,1.00}  \\\cline{1-7}
\end{tabular}
\caption{Retractive index of the i-th layer of the media used in the reported 3D simulations. The layer-0 is the external environment (see Fig.~\ref{lsphereA}). In the media D and E the refractive index  varies linearly in the 100 layers between the two extreme values by an increment of 0.01. }
\label{tab}
\end{table}

The isotropic uniform illumination condition has been implemented by  reproducing a Lambertian distribution  of the entrance angles, that is a distribution that follows the cosine law. A crossover of an interface between two different media is considered as a new scattering event in term of step length extraction, whereas the change in direction is deterministically established by the Snell law.

Also the radiance $I$ and the fluence $\Phi$ (the integral over the solid angle of the radiance $I$) have been calculated, considering the external surface of the medium as illuminated by a unitary input flux of 1 W/m$^2$ (see the Appendix \ref{appendix}). 

Then, within a generic subvolume $V_i$ of the medium, the radiance is constant and equal to:
\begin{equation}
I_i=\frac{1}{\pi}\left(n_i/n_e\right)^2\,\text{W/m$^2$sr} 
\label{total_flux4}
\end{equation}
The corresponding fluence is as follows:
\begin{equation}
\Phi_i=4\pi I_i= 4 \left(n_i/n_e\right)^2\,\text{W/m$^2$} 
\label{total_flux5}
\end{equation}

An independent MC code has been used for the 2D case, considering the same expression (Eq.~\ref{LB2}) for the generation of the length between two scattering events. Its robustness has been then tested with the analytical results for the IP in condition of absence of refractive index mismatches.

In Fig.~\ref{es_rw_2D}, a simulated trajectory is shown, considering a 2D circular medium with a circular de-centered inhomogeneity with different refractive indices and  scattering coefficients. Also, a refractive-index mismatch is present between the external environment and the largest medium.  

\section{Results}\label{sim_risultati}
In all the results shown in this paper, the label ``$SIM$'' is referred to the data generated by MC simulations, whereas the label ``$IP$'' is referred to values obtained by the theory. 
\subsection{The 3D-case}

In Fig.~\ref{lsphereA} the structure of a medium simulated in the 3D case is shown.  A sphere composed of four layers, characterized by a refractive index $n_i$ and a scattering coefficient $\mu_i$,  is immersed in an external environment with a refractive index $n_0$. The radiation impinges upon the external layer from the outside uniformly and  isotropically.
Five different cases, characterized by different parameters, are labeled with the letters $A$ to $E$ and described in Tab.~\ref{tab} (superscript in the plot legend).  The external environment is labeled as Layer-0. The radius of the sphere is 5 mm and, in the cases  $A$, $B$ and $C$, the second, the third and the fourth layer have radius 4, 3 and 2 mm respectively. The cases $D$ and $E$ are characterized by 100 layers, with a refractive index that becomes larger or smaller as the depth $d$ from the outermost layer increases.

For the same $\mu_s$  in the whole medium, each MC calculation consists of 100 independent simulations of $10^8$ (case $A$ to $C$) or $10^7$ (case $D$ and $E$) independent trajectories,  in order to evaluate the standard error of the averaged mean path length $\langle L\rangle_{SIM}$.   

\begin{figure}
\centerline{{\includegraphics[width=1.0\columnwidth]{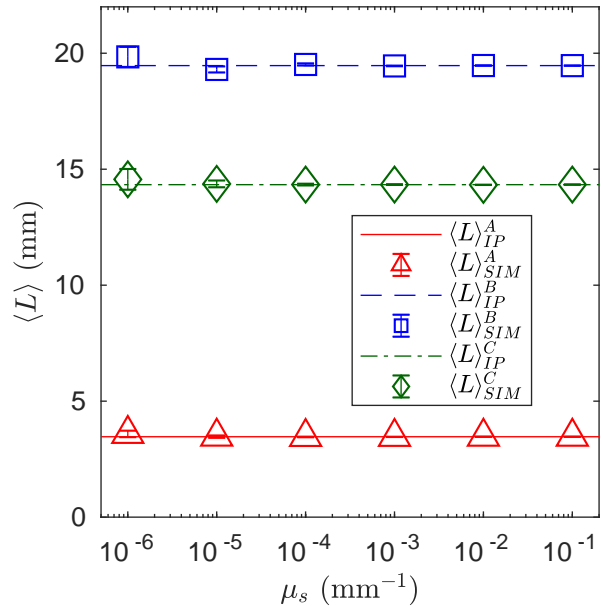}}}
\caption{(color online) The 3D-case: $\langle L\rangle$ for different values of $\mu_s$ for the media A, B and C (see Tab.~\ref{tab}). Markers are referred to MC simulations and lines to the values expected by IP. }
\label{fig_rw_3D_AB}
\end{figure}

Figure \ref{fig_rw_3D_AB} shows $\langle L\rangle_{SIM}$ as a function of $\mu_s$ for the media A,B and C. In these cases the strong refractive index mismatch between two adjacent layers leads to a relatively high reflection at the interface.
In the figure, the values predicted by the generalized IP (Eq.~(\ref{pALBInvarianceInhom4}))  are also reported. The results show that the mean path length of the random walkers is consistent with the value predicted by theory over five orders of magnitude of the scattering coefficient $\mu_s$.
  
\begin{figure}
\centerline{{\includegraphics[width=1.0\columnwidth]{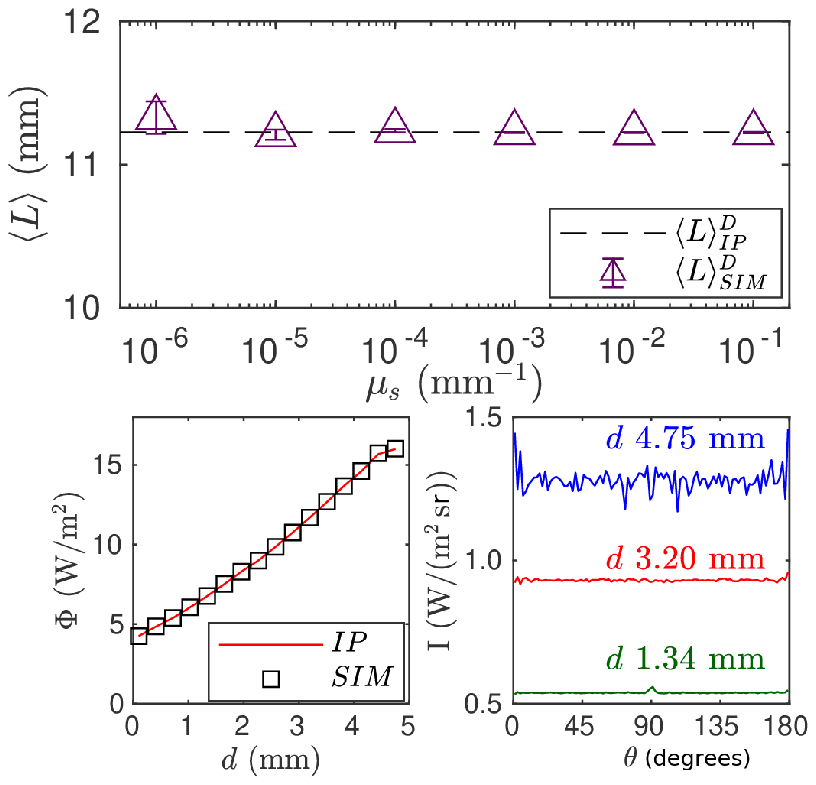}}}
\caption{(color online) The 3D-case: on the top,  $\langle L\rangle$ for  different vale of $\mu_s$ for the medium D (see Tab.~\ref{tab}). Markers are referred to MC simulations and lines to the values expected by IP. For the case of $\mu_s=10^{-1}$ mm$^{-1}$, on bottom-left and on bottom-right, the fluence (the markers are MC data and lines are IP values) as a function of the depth $d$ from the surface of the largest sphere and the radiance data of MC simulations for three depths are respectively reported.   }
\label{fig_rw_3D_E}
\end{figure}
\begin{figure}
\centerline{{\includegraphics[width=1.0\columnwidth]{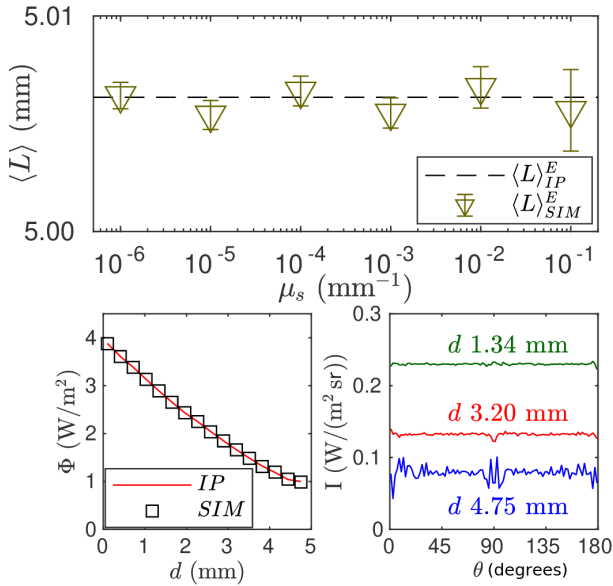}}}
\caption{(color online) The 3D-case: on the top,  the mean path-length $\langle L\rangle$ for different values of $\mu_s$ for the medium E (see Tab.~\ref{tab}). Markers are referred to MC simulations and lines to the values expected by IP. For the case of $\mu_s=10^{-1}$ mm$^{-1}$, on bottom-left and on bottom-right, the fluence (the markers are MC data and lines are IP values) as a function of the depth $d$ from the surface of the largest sphere and the radiance data of MC simulations for three depths are reported  respectively .   }
\label{fig_rw_3D_F}
\end{figure}

Figures \ref{fig_rw_3D_E} and \ref{fig_rw_3D_F} show the results for the media composed by 100 layers with a refractive index that linearly increases or decreases as the layer index $i$ grows (see Tab.~\ref{tab}), or, equivalently, as the distance $d$ from the surface increases.  In these figures, the fluence as a function of $d$ and the profile of the radiance for three values of $d$ are also shown. The fluence has a constant value within each layer and increasing ordecreasing values as the layer change. The radiance, within the noise of the MC simulations, shows a constant angular profile, as expected where the IP holds.

\subsection{The 2D-case}
In the simulations of the 2D case,  we analyze the cases of homogeneous and non-homogeneous media. As shown in Fig.~\ref{es_rw_2D}, a typical medium consists of a circle immersed in an external environment and with an inhomogeneity inside. Such an inhomogeneity is a circle with a different refractive index and scattering coefficient and has a center located  at a random position respect to the center of the larger circle. Each result, for a fixed set of parameters, consists of 10 independent simulations of 2$\cdot$10$^8$ trajectories.

\begin{figure}
\centerline{{\includegraphics[width=1.0\columnwidth]{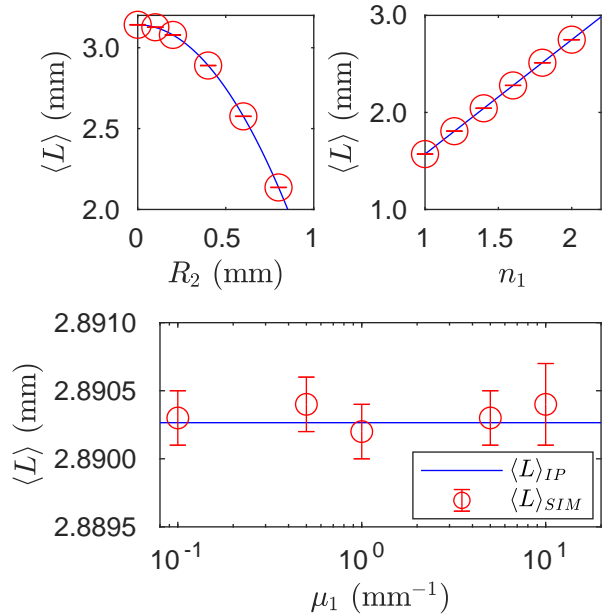}}}
\caption{(color online) 2D-case: $\langle L\rangle$ for different values of the radius $R_2$ of the circular de-centered inhomogeneity (top-left figure), of the refractive index $n_1$ (top-right figure) of the scattering coefficient $\mu_1$ of the largest medium (bottom figure) (Markers are MC data and lines are IP values).
}
\label{fig_sim_2D}
\end{figure}

In Fig.~\ref{fig_sim_2D} (top-left), the mean path-length $\langle L\rangle$ for different values of the radius $R_2$ of the circular decentered inhomogeneity is shown. The largest medium has radius $R_1=1$ mm.
The refractive indexes of the external environment ($n_0$), the largest medium ($n_1$) and the inhomogeneity ($n_2$) are 1, 2 and 1 respectively.  The scattering coefficient is 0.5 mm$^{-1}$ everywhere. 

Fig.~\ref{fig_sim_2D} (top-right) shows $\langle L\rangle$ for different values of the refractive index $n_1$ of the largest medium. The refractive index is 1 in the external environment and in the de-centered inhomogeneity. The radii of the largest medium and of the inhomogeneity are $R_1=1$ mm and $R_2=0.5$ mm respectively.
The scattering coefficient is everywhere constant (0.5 mm$^{-1}$).

In Fig.~\ref{fig_sim_2D} (bottom), the refractive index of the external environment ($n_0$) and of the de-centered inhomogeneity $n_2$ are 1, whereas $n_1$ is 2. The radii of the largest medium and of the inhomogeneity is $R_1=1$ mm and $R_2=0.4$ mm respectively, while  the scattering coefficient of the inhomogeneity is 1 mm$^{-1}$. In this case,   $\langle L\rangle$ is shown as a function of the scattering coefficient $\mu_1$ of the largest medium.

In all the 2D cases analyzed, the MC results for $\langle L\rangle$ (red markers) are consistent within the standard error (of a order of $10^{-3}\div 10^{-4}$) with the value predicted by the Eq.~(\ref{pALBInvarianceInhom4surface}) (solid blue line).

\section{Conclusions}
The mean path length invariance property has been investigated in the case  of light diffusion in the presence of refractive index mismatch between a medium and the external environment as well as for inhomogeneous media. Analytical results are found both for the two dimensional and the three dimensional cases. Numerical calculations are in full agreement with analytical expressions. 

Thus, in the paper the invariance property is shown to remain valid for all  classes of inhomogeneous media with discrete variations of refractive index in their internal domain.
This is a further generalization of this property that also opens the possibility to use its prediction to a quite larger class of problems. This fact can be exploited to obtain reference values for photon migration studies in random media. For instance, the IP represents a unique and powerful tool able to validate any MC code extensively used to simulate random walk processes with an arbitrary precision.

From the point of view of practical applications in tissue optics and photovoltaics, such a paper helps to understand the optical propagation in the most general kind of medium, i.e.\ an inhomogeneous material with scattering and refractive index mismatches.

\section*{Acknowledgments}
Acknowledgments are due to professor G.~ Zaccanti for his help in performing the MC simulations. 

\appendix
\section{Total input flux}\label{appendix}
The shown results for the radiance and the fluence are referred to an external surface illuminated by a uniform isotropic illumination for the radiance (Lambetian illumination) with the further hypothesis to have a unitary  total input flux  in the medium, given by the integral of the radiance over the whole solid semi-angle \cite{librofabrizio}. This fact implies that only the normal component of $\vec{J}_{in}$ is not null. Given $I_{in}$ as the input radiance, we have that
\begin{equation}
J_{in}=\int\limits_{2\pi}I_{in}\hat{s}\cdot\hat{n}d\hat{s}=\pi I_{in}
\end{equation}
and, then,
\begin{equation}
I_{in}=\frac{1\,\text{W/m$^2$}}{\pi}.
\end{equation}
The corresponding input fluence is as follows:
\begin{equation}
\Phi_{in}=\int\limits_{2\pi}I_{in}d\hat{s}=2\,\text{W/m$^2$}
\end{equation}
In a generic subvolume, the radiance is as follows:
\begin{equation}
I_i=\left( \frac{n_i}{n_e} \right)^2I_{in}
\end{equation}
and the fluence,
\begin{equation}
\Phi_i=4\pi I_i
\end{equation}


\begin{thebibliography}{45}%
\makeatletter
\providecommand \@ifxundefined [1]{%
 \@ifx{#1\undefined}
}%
\providecommand \@ifnum [1]{%
 \ifnum #1\expandafter \@firstoftwo
 \else \expandafter \@secondoftwo
 \fi
}%
\providecommand \@ifx [1]{%
 \ifx #1\expandafter \@firstoftwo
 \else \expandafter \@secondoftwo
 \fi
}%
\providecommand \natexlab [1]{#1}%
\providecommand \enquote  [1]{``#1''}%
\providecommand \bibnamefont  [1]{#1}%
\providecommand \bibfnamefont [1]{#1}%
\providecommand \citenamefont [1]{#1}%
\providecommand \href@noop [0]{\@secondoftwo}%
\providecommand \href [0]{\begingroup \@sanitize@url \@href}%
\providecommand \@href[1]{\@@startlink{#1}\@@href}%
\providecommand \@@href[1]{\endgroup#1\@@endlink}%
\providecommand \@sanitize@url [0]{\catcode `\\12\catcode `\$12\catcode
  `\&12\catcode `\#12\catcode `\^12\catcode `\_12\catcode `\%12\relax}%
\providecommand \@@startlink[1]{}%
\providecommand \@@endlink[0]{}%
\providecommand \url  [0]{\begingroup\@sanitize@url \@url }%
\providecommand \@url [1]{\endgroup\@href {#1}{\urlprefix }}%
\providecommand \urlprefix  [0]{URL }%
\providecommand \Eprint [0]{\href }%
\providecommand \doibase [0]{http://dx.doi.org/}%
\providecommand \selectlanguage [0]{\@gobble}%
\providecommand \bibinfo  [0]{\@secondoftwo}%
\providecommand \bibfield  [0]{\@secondoftwo}%
\providecommand \translation [1]{[#1]}%
\providecommand \BibitemOpen [0]{}%
\providecommand \bibitemStop [0]{}%
\providecommand \bibitemNoStop [0]{.\EOS\space}%
\providecommand \EOS [0]{\spacefactor3000\relax}%
\providecommand \BibitemShut  [1]{\csname bibitem#1\endcsname}%
\let\auto@bib@innerbib\@empty
\bibitem [{\citenamefont {Jacques}(2013)}]{Jacques_2013}%
  \BibitemOpen
  \bibfield  {author} {\bibinfo {author} {\bibfnamefont {S.~L.}\ \bibnamefont
  {Jacques}},\ }\href {\doibase 10.1088/0031-9155/58/11/r37} {\bibfield
  {journal} {\bibinfo  {journal} {Phys. Med. Biol.}\ }\textbf {\bibinfo
  {volume} {58}},\ \bibinfo {pages} {R37} (\bibinfo {year} {2013})}\BibitemShut
  {NoStop}%
\bibitem [{\citenamefont {Durduran}\ \emph {et~al.}(2010)\citenamefont
  {Durduran}, \citenamefont {Choe}, \citenamefont {Baker},\ and\ \citenamefont
  {Yodh}}]{Durduran_2010}%
  \BibitemOpen
  \bibfield  {author} {\bibinfo {author} {\bibfnamefont {T.}~\bibnamefont
  {Durduran}}, \bibinfo {author} {\bibfnamefont {R.}~\bibnamefont {Choe}},
  \bibinfo {author} {\bibfnamefont {W.~B.}\ \bibnamefont {Baker}}, \ and\
  \bibinfo {author} {\bibfnamefont {A.~G.}\ \bibnamefont {Yodh}},\ }\href
  {\doibase 10.1088/0034-4885/73/7/076701} {\bibfield  {journal} {\bibinfo
  {journal} {Reports on Progress in Physics}\ }\textbf {\bibinfo {volume}
  {73}},\ \bibinfo {pages} {076701} (\bibinfo {year} {2010})}\BibitemShut
  {NoStop}%
\bibitem [{\citenamefont {Leff}\ \emph {et~al.}(2008)\citenamefont {Leff},
  \citenamefont {Warren}, \citenamefont {Enfield}, \citenamefont {Gibson},
  \citenamefont {Athanasiou}, \citenamefont {Patten}, \citenamefont {Hebden},
  \citenamefont {Yang},\ and\ \citenamefont {Darzi}}]{Leff2008}%
  \BibitemOpen
  \bibfield  {author} {\bibinfo {author} {\bibfnamefont {D.~R.}\ \bibnamefont
  {Leff}}, \bibinfo {author} {\bibfnamefont {O.~J.}\ \bibnamefont {Warren}},
  \bibinfo {author} {\bibfnamefont {L.~C.}\ \bibnamefont {Enfield}}, \bibinfo
  {author} {\bibfnamefont {A.}~\bibnamefont {Gibson}}, \bibinfo {author}
  {\bibfnamefont {T.}~\bibnamefont {Athanasiou}}, \bibinfo {author}
  {\bibfnamefont {D.~K.}\ \bibnamefont {Patten}}, \bibinfo {author}
  {\bibfnamefont {J.}~\bibnamefont {Hebden}}, \bibinfo {author} {\bibfnamefont
  {G.~Z.}\ \bibnamefont {Yang}}, \ and\ \bibinfo {author} {\bibfnamefont
  {A.}~\bibnamefont {Darzi}},\ }\href {\doibase 10.1007/s10549-007-9582-z}
  {\bibfield  {journal} {\bibinfo  {journal} {Breast Cancer Research and
  Treatment}\ }\textbf {\bibinfo {volume} {108}},\ \bibinfo {pages} {9}
  (\bibinfo {year} {2008})}\BibitemShut {NoStop}%
\bibitem [{\citenamefont {Zaccanti}\ \emph {et~al.}(2003)\citenamefont
  {Zaccanti}, \citenamefont {Bianco},\ and\ \citenamefont
  {Martelli}}]{Zaccanti:03}%
  \BibitemOpen
  \bibfield  {author} {\bibinfo {author} {\bibfnamefont {G.}~\bibnamefont
  {Zaccanti}}, \bibinfo {author} {\bibfnamefont {S.~D.}\ \bibnamefont
  {Bianco}}, \ and\ \bibinfo {author} {\bibfnamefont {F.}~\bibnamefont
  {Martelli}},\ }\href {\doibase 10.1364/AO.42.004023} {\bibfield  {journal}
  {\bibinfo  {journal} {Appl. Opt.}\ }\textbf {\bibinfo {volume} {42}},\
  \bibinfo {pages} {4023} (\bibinfo {year} {2003})}\BibitemShut {NoStop}%
\bibitem [{\citenamefont {Wiersma}(2008)}]{rl2}%
  \BibitemOpen
  \bibfield  {author} {\bibinfo {author} {\bibfnamefont {D.~S.}\ \bibnamefont
  {Wiersma}},\ }\href@noop {} {\bibfield  {journal} {\bibinfo  {journal} {Nat.
  Phys.}\ }\textbf {\bibinfo {volume} {4}},\ \bibinfo {pages} {359} (\bibinfo
  {year} {2008})}\BibitemShut {NoStop}%
\bibitem [{\citenamefont {Lawandy}\ \emph {et~al.}(1995)\citenamefont
  {Lawandy}, \citenamefont {Balachandran}, \citenamefont {Gomes},\ and\
  \citenamefont {Suvain}}]{rl3}%
  \BibitemOpen
  \bibfield  {author} {\bibinfo {author} {\bibfnamefont {N.~M.}\ \bibnamefont
  {Lawandy}}, \bibinfo {author} {\bibfnamefont {R.~M.}\ \bibnamefont
  {Balachandran}}, \bibinfo {author} {\bibfnamefont {A.~S.~L.}\ \bibnamefont
  {Gomes}}, \ and\ \bibinfo {author} {\bibfnamefont {E.}~\bibnamefont
  {Suvain}},\ }\href@noop {} {\bibfield  {journal} {\bibinfo  {journal}
  {Nature}\ }\textbf {\bibinfo {volume} {368}},\ \bibinfo {pages} {436}
  (\bibinfo {year} {1995})}\BibitemShut {NoStop}%
\bibitem [{\citenamefont {Ignesti}\ \emph {et~al.}(2013)\citenamefont
  {Ignesti}, \citenamefont {Tommasi}, \citenamefont {Fini}, \citenamefont
  {Lepri}, \citenamefont {Radhalakshmi}, \citenamefont {Wiersma},\ and\
  \citenamefont {Cavalieri}}]{nostro1}%
  \BibitemOpen
  \bibfield  {author} {\bibinfo {author} {\bibfnamefont {E.}~\bibnamefont
  {Ignesti}}, \bibinfo {author} {\bibfnamefont {F.}~\bibnamefont {Tommasi}},
  \bibinfo {author} {\bibfnamefont {L.}~\bibnamefont {Fini}}, \bibinfo {author}
  {\bibfnamefont {S.}~\bibnamefont {Lepri}}, \bibinfo {author} {\bibfnamefont
  {V.}~\bibnamefont {Radhalakshmi}}, \bibinfo {author} {\bibfnamefont {D.~S.}\
  \bibnamefont {Wiersma}}, \ and\ \bibinfo {author} {\bibfnamefont
  {S.}~\bibnamefont {Cavalieri}},\ }\href@noop {} {\bibfield  {journal}
  {\bibinfo  {journal} {Phys. Rev. A}\ }\textbf {\bibinfo {volume} {88}},\
  \bibinfo {pages} {033820} (\bibinfo {year} {2013})}\BibitemShut {NoStop}%
\bibitem [{\citenamefont {Tommasi}\ \emph {et~al.}(2015)\citenamefont
  {Tommasi}, \citenamefont {Ignesti}, \citenamefont {Fini},\ and\ \citenamefont
  {Cavalieri}}]{PhysRevA.91.033820}%
  \BibitemOpen
  \bibfield  {author} {\bibinfo {author} {\bibfnamefont {F.}~\bibnamefont
  {Tommasi}}, \bibinfo {author} {\bibfnamefont {E.}~\bibnamefont {Ignesti}},
  \bibinfo {author} {\bibfnamefont {L.}~\bibnamefont {Fini}}, \ and\ \bibinfo
  {author} {\bibfnamefont {S.}~\bibnamefont {Cavalieri}},\ }\href {\doibase
  10.1103/PhysRevA.91.033820} {\bibfield  {journal} {\bibinfo  {journal} {Phys.
  Rev. A}\ }\textbf {\bibinfo {volume} {91}},\ \bibinfo {pages} {033820}
  (\bibinfo {year} {2015})}\BibitemShut {NoStop}%
\bibitem [{\citenamefont {Tommasi}\ \emph
  {et~al.}(2018{\natexlab{a}})\citenamefont {Tommasi}, \citenamefont {Fini},
  \citenamefont {Ignesti}, \citenamefont {Lepri}, \citenamefont {Martelli},\
  and\ \citenamefont {Cavalieri}}]{PhysRevA.98.053816}%
  \BibitemOpen
  \bibfield  {author} {\bibinfo {author} {\bibfnamefont {F.}~\bibnamefont
  {Tommasi}}, \bibinfo {author} {\bibfnamefont {L.}~\bibnamefont {Fini}},
  \bibinfo {author} {\bibfnamefont {E.}~\bibnamefont {Ignesti}}, \bibinfo
  {author} {\bibfnamefont {S.}~\bibnamefont {Lepri}}, \bibinfo {author}
  {\bibfnamefont {F.}~\bibnamefont {Martelli}}, \ and\ \bibinfo {author}
  {\bibfnamefont {S.}~\bibnamefont {Cavalieri}},\ }\href@noop {} {\bibfield
  {journal} {\bibinfo  {journal} {Phys. Rev. A}\ }\textbf {\bibinfo {volume}
  {98}},\ \bibinfo {pages} {053816} (\bibinfo {year}
  {2018}{\natexlab{a}})}\BibitemShut {NoStop}%
\bibitem [{\citenamefont {Segev}\ \emph {et~al.}(2013)\citenamefont {Segev},
  \citenamefont {Silberberg},\ and\ \citenamefont {Christodoulides}}]{and_loc}%
  \BibitemOpen
  \bibfield  {author} {\bibinfo {author} {\bibfnamefont {M.}~\bibnamefont
  {Segev}}, \bibinfo {author} {\bibfnamefont {Y.}~\bibnamefont {Silberberg}}, \
  and\ \bibinfo {author} {\bibfnamefont {D.~N.}\ \bibnamefont
  {Christodoulides}},\ }\href@noop {} {\bibfield  {journal} {\bibinfo
  {journal} {Nat. Photonics}\ }\textbf {\bibinfo {volume} {7}},\ \bibinfo
  {pages} {197} (\bibinfo {year} {2013})}\BibitemShut {NoStop}%
\bibitem [{\citenamefont {Cao}\ \emph {et~al.}(2000)\citenamefont {Cao},
  \citenamefont {Xu}, \citenamefont {Zhang}, \citenamefont {Ho}, \citenamefont
  {Seelig}, \citenamefont {Liu},\ and\ \citenamefont {Chang}}]{and1}%
  \BibitemOpen
  \bibfield  {author} {\bibinfo {author} {\bibfnamefont {H.}~\bibnamefont
  {Cao}}, \bibinfo {author} {\bibfnamefont {J.~Y.}\ \bibnamefont {Xu}},
  \bibinfo {author} {\bibfnamefont {D.~Z.}\ \bibnamefont {Zhang}, \bibfnamefont
  {S.~H.~Chang}}, \bibinfo {author} {\bibfnamefont {S.~T.}\ \bibnamefont {Ho}},
  \bibinfo {author} {\bibfnamefont {E.~W.}\ \bibnamefont {Seelig}}, \bibinfo
  {author} {\bibfnamefont {X.}~\bibnamefont {Liu}}, \ and\ \bibinfo {author}
  {\bibfnamefont {R.~P.~H.}\ \bibnamefont {Chang}},\ }\href@noop {} {\bibfield
  {journal} {\bibinfo  {journal} {Phys. Rev. Lett.}\ }\textbf {\bibinfo
  {volume} {84(\normalfont{24})}},\ \bibinfo {pages} {5584} (\bibinfo {year}
  {2000})}\BibitemShut {NoStop}%
\bibitem [{\citenamefont {Barthelemy}\ \emph {et~al.}(2008)\citenamefont
  {Barthelemy}, \citenamefont {Bertolotti},\ and\ \citenamefont
  {Wiersma}}]{lf1}%
  \BibitemOpen
  \bibfield  {author} {\bibinfo {author} {\bibfnamefont {P.}~\bibnamefont
  {Barthelemy}}, \bibinfo {author} {\bibfnamefont {J.}~\bibnamefont
  {Bertolotti}}, \ and\ \bibinfo {author} {\bibfnamefont {D.~S.}\ \bibnamefont
  {Wiersma}},\ }\href@noop {} {\bibfield  {journal} {\bibinfo  {journal}
  {Nature}\ }\textbf {\bibinfo {volume} {453}},\ \bibinfo {pages} {459}
  (\bibinfo {year} {2008})}\BibitemShut {NoStop}%
\bibitem [{\citenamefont {Bertolotti}\ \emph {et~al.}(2010)\citenamefont
  {Bertolotti}, \citenamefont {Vynck},\ and\ \citenamefont {Wiersma}}]{LF2}%
  \BibitemOpen
  \bibfield  {author} {\bibinfo {author} {\bibfnamefont {J.}~\bibnamefont
  {Bertolotti}}, \bibinfo {author} {\bibfnamefont {K.}~\bibnamefont {Vynck}}, \
  and\ \bibinfo {author} {\bibfnamefont {D.~S.}\ \bibnamefont {Wiersma}},\
  }\href@noop {} {\bibfield  {journal} {\bibinfo  {journal} {Phys. Rev. Lett.}\
  }\textbf {\bibinfo {volume} {105}},\ \bibinfo {pages} {163902} (\bibinfo
  {year} {2010})}\BibitemShut {NoStop}%
\bibitem [{\citenamefont {Tommasi}\ \emph {et~al.}(2019)\citenamefont
  {Tommasi}, \citenamefont {Fini}, \citenamefont {Martelli},\ and\
  \citenamefont {Cavalieri}}]{PhysRevA.99.063836}%
  \BibitemOpen
  \bibfield  {author} {\bibinfo {author} {\bibfnamefont {F.}~\bibnamefont
  {Tommasi}}, \bibinfo {author} {\bibfnamefont {L.}~\bibnamefont {Fini}},
  \bibinfo {author} {\bibfnamefont {F.}~\bibnamefont {Martelli}}, \ and\
  \bibinfo {author} {\bibfnamefont {S.}~\bibnamefont {Cavalieri}},\ }\href
  {\doibase 10.1103/PhysRevA.99.063836} {\bibfield  {journal} {\bibinfo
  {journal} {Phys. Rev. A}\ }\textbf {\bibinfo {volume} {99}},\ \bibinfo
  {pages} {063836} (\bibinfo {year} {2019})}\BibitemShut {NoStop}%
\bibitem [{\citenamefont {Gofraniha}\ \emph {et~al.}(2015)\citenamefont
  {Gofraniha}, \citenamefont {Viola}, \citenamefont {Di~Maria}, \citenamefont
  {Barbarella}, \citenamefont {Gigli}, \citenamefont {Leuzzi},\ and\
  \citenamefont {Conti}}]{rsb1}%
  \BibitemOpen
  \bibfield  {author} {\bibinfo {author} {\bibfnamefont {N.}~\bibnamefont
  {Gofraniha}}, \bibinfo {author} {\bibfnamefont {I.}~\bibnamefont {Viola}},
  \bibinfo {author} {\bibfnamefont {F.}~\bibnamefont {Di~Maria}}, \bibinfo
  {author} {\bibfnamefont {G.}~\bibnamefont {Barbarella}}, \bibinfo {author}
  {\bibfnamefont {G.}~\bibnamefont {Gigli}}, \bibinfo {author} {\bibfnamefont
  {L.}~\bibnamefont {Leuzzi}}, \ and\ \bibinfo {author} {\bibfnamefont
  {C.}~\bibnamefont {Conti}},\ }\href@noop {} {\bibfield  {journal} {\bibinfo
  {journal} {Nat. Commun.}\ }\textbf {\bibinfo {volume} {6}},\ \bibinfo {pages}
  {6058} (\bibinfo {year} {2015})}\BibitemShut {NoStop}%
\bibitem [{\citenamefont {Gomes}\ \emph {et~al.}(2016)\citenamefont {Gomes},
  \citenamefont {Raposo}, \citenamefont {Moura}, \citenamefont {Fewo},
  \citenamefont {Pincheira}, \citenamefont {Jerez},\ and\ \citenamefont
  {Maia}}]{rsb3}%
  \BibitemOpen
  \bibfield  {author} {\bibinfo {author} {\bibfnamefont {A.~S.~L.}\
  \bibnamefont {Gomes}}, \bibinfo {author} {\bibfnamefont {E.~P.}\ \bibnamefont
  {Raposo}}, \bibinfo {author} {\bibfnamefont {A.~L.}\ \bibnamefont {Moura}},
  \bibinfo {author} {\bibfnamefont {S.~I.}\ \bibnamefont {Fewo}}, \bibinfo
  {author} {\bibfnamefont {P.~I.~R.}\ \bibnamefont {Pincheira}}, \bibinfo
  {author} {\bibfnamefont {V.}~\bibnamefont {Jerez}}, \ and\ \bibinfo {author}
  {\bibfnamefont {L.~J. Q. d.~A.}\ \bibnamefont {Maia}},\ }\href@noop {}
  {\bibfield  {journal} {\bibinfo  {journal} {Sci. Rep.}\ }\textbf {\bibinfo
  {volume} {5}},\ \bibinfo {pages} {27987} (\bibinfo {year}
  {2016})}\BibitemShut {NoStop}%
\bibitem [{\citenamefont {Tommasi}\ \emph {et~al.}(2016)\citenamefont
  {Tommasi}, \citenamefont {Ignesti}, \citenamefont {Lepri},\ and\
  \citenamefont {Cavalieri}}]{rsb4}%
  \BibitemOpen
  \bibfield  {author} {\bibinfo {author} {\bibfnamefont {F.}~\bibnamefont
  {Tommasi}}, \bibinfo {author} {\bibfnamefont {E.}~\bibnamefont {Ignesti}},
  \bibinfo {author} {\bibfnamefont {S.}~\bibnamefont {Lepri}}, \ and\ \bibinfo
  {author} {\bibfnamefont {S.}~\bibnamefont {Cavalieri}},\ }\href@noop {}
  {\bibfield  {journal} {\bibinfo  {journal} {Sci. Rep.}\ }\textbf {\bibinfo
  {volume} {6}},\ \bibinfo {pages} {37113} (\bibinfo {year}
  {2016})}\BibitemShut {NoStop}%
\bibitem [{\citenamefont {Polson}\ and\ \citenamefont
  {Vardeny}(2004)}]{doi:10.1063/1.1782259}%
  \BibitemOpen
  \bibfield  {author} {\bibinfo {author} {\bibfnamefont {R.~C.}\ \bibnamefont
  {Polson}}\ and\ \bibinfo {author} {\bibfnamefont {Z.~V.}\ \bibnamefont
  {Vardeny}},\ }\href {\doibase 10.1063/1.1782259} {\bibfield  {journal}
  {\bibinfo  {journal} {Appl. Phys. Lett.}\ }\textbf {\bibinfo {volume} {85}},\
  \bibinfo {pages} {1289} (\bibinfo {year} {2004})}\BibitemShut {NoStop}%
\bibitem [{\citenamefont {{Di Ninni}}\ \emph {et~al.}(2011)\citenamefont {{Di
  Ninni}}, \citenamefont {Martelli},\ and\ \citenamefont
  {Zaccanti}}]{0031-9155-56-2-N01}%
  \BibitemOpen
  \bibfield  {author} {\bibinfo {author} {\bibfnamefont {P.}~\bibnamefont {{Di
  Ninni}}}, \bibinfo {author} {\bibfnamefont {F.}~\bibnamefont {Martelli}}, \
  and\ \bibinfo {author} {\bibfnamefont {G.}~\bibnamefont {Zaccanti}},\
  }\href@noop {} {\bibfield  {journal} {\bibinfo  {journal} {Phys. Med. Biol.}\
  }\textbf {\bibinfo {volume} {56(\normalfont{2})}},\ \bibinfo {pages} {N21}
  (\bibinfo {year} {2011})}\BibitemShut {NoStop}%
\bibitem [{\citenamefont {Ignesti}\ \emph {et~al.}(2016)\citenamefont
  {Ignesti}, \citenamefont {Tommasi}, \citenamefont {Fini}, \citenamefont
  {Martelli}, \citenamefont {Azzali},\ and\ \citenamefont
  {Cavalieri}}]{sensore}%
  \BibitemOpen
  \bibfield  {author} {\bibinfo {author} {\bibfnamefont {E.}~\bibnamefont
  {Ignesti}}, \bibinfo {author} {\bibfnamefont {F.}~\bibnamefont {Tommasi}},
  \bibinfo {author} {\bibfnamefont {L.}~\bibnamefont {Fini}}, \bibinfo {author}
  {\bibfnamefont {F.}~\bibnamefont {Martelli}}, \bibinfo {author}
  {\bibfnamefont {N.}~\bibnamefont {Azzali}}, \ and\ \bibinfo {author}
  {\bibfnamefont {S.}~\bibnamefont {Cavalieri}},\ }\href@noop {} {\bibfield
  {journal} {\bibinfo  {journal} {Sci. Rep.}\ }\textbf {\bibinfo {volume}
  {6}},\ \bibinfo {pages} {35225} (\bibinfo {year} {2016})}\BibitemShut
  {NoStop}%
\bibitem [{\citenamefont {Vasquez}\ \emph {et~al.}(2018)\citenamefont
  {Vasquez}, \citenamefont {Hern\`andez},\ and\ \citenamefont
  {Coello}}]{sens_scat}%
  \BibitemOpen
  \bibfield  {author} {\bibinfo {author} {\bibfnamefont {G.}~\bibnamefont
  {Vasquez}}, \bibinfo {author} {\bibfnamefont {Y.}~\bibnamefont
  {Hern\`andez}}, \ and\ \bibinfo {author} {\bibfnamefont {Y.}~\bibnamefont
  {Coello}},\ }\href@noop {} {\bibfield  {journal} {\bibinfo  {journal} {Sci.
  Rep.}\ }\textbf {\bibinfo {volume} {8}},\ \bibinfo {pages} {14903} (\bibinfo
  {year} {2018})}\BibitemShut {NoStop}%
\bibitem [{\citenamefont {Tommasi}\ \emph
  {et~al.}(2018{\natexlab{b}})\citenamefont {Tommasi}, \citenamefont {Ignesti},
  \citenamefont {Fini}, \citenamefont {Martelli},\ and\ \citenamefont
  {Cavalieri}}]{Tommasi:18}%
  \BibitemOpen
  \bibfield  {author} {\bibinfo {author} {\bibfnamefont {F.}~\bibnamefont
  {Tommasi}}, \bibinfo {author} {\bibfnamefont {E.}~\bibnamefont {Ignesti}},
  \bibinfo {author} {\bibfnamefont {L.}~\bibnamefont {Fini}}, \bibinfo {author}
  {\bibfnamefont {F.}~\bibnamefont {Martelli}}, \ and\ \bibinfo {author}
  {\bibfnamefont {S.}~\bibnamefont {Cavalieri}},\ }\href@noop {} {\bibfield
  {journal} {\bibinfo  {journal} {Opt. Express}\ }\textbf {\bibinfo {volume}
  {26}},\ \bibinfo {pages} {27615} (\bibinfo {year}
  {2018}{\natexlab{b}})}\BibitemShut {NoStop}%
\bibitem [{\citenamefont {Atwater}\ and\ \citenamefont
  {Polman}(2010)}]{atwater}%
  \BibitemOpen
  \bibfield  {author} {\bibinfo {author} {\bibfnamefont {H.~A.}\ \bibnamefont
  {Atwater}}\ and\ \bibinfo {author} {\bibfnamefont {A.}~\bibnamefont
  {Polman}},\ }\href@noop {} {\bibfield  {journal} {\bibinfo  {journal} {Nat.
  Mater.}\ }\textbf {\bibinfo {volume} {9}},\ \bibinfo {pages} {205} (\bibinfo
  {year} {2010})}\BibitemShut {NoStop}%
\bibitem [{\citenamefont {Green}\ and\ \citenamefont {Pillai}(2012)}]{green}%
  \BibitemOpen
  \bibfield  {author} {\bibinfo {author} {\bibfnamefont {M.~A.}\ \bibnamefont
  {Green}}\ and\ \bibinfo {author} {\bibfnamefont {S.}~\bibnamefont {Pillai}},\
  }\href@noop {} {\bibfield  {journal} {\bibinfo  {journal} {Nat. Photonics}\
  }\textbf {\bibinfo {volume} {6}},\ \bibinfo {pages} {130} (\bibinfo {year}
  {2012})}\BibitemShut {NoStop}%
\bibitem [{\citenamefont {Mupparapu}\ \emph {et~al.}(2015)\citenamefont
  {Mupparapu}, \citenamefont {Vynck}, \citenamefont {Svensson}, \citenamefont
  {Burresi},\ and\ \citenamefont {Wiersma}}]{Mupparapu:15}%
  \BibitemOpen
  \bibfield  {author} {\bibinfo {author} {\bibfnamefont {R.}~\bibnamefont
  {Mupparapu}}, \bibinfo {author} {\bibfnamefont {K.}~\bibnamefont {Vynck}},
  \bibinfo {author} {\bibfnamefont {T.}~\bibnamefont {Svensson}}, \bibinfo
  {author} {\bibfnamefont {M.}~\bibnamefont {Burresi}}, \ and\ \bibinfo
  {author} {\bibfnamefont {D.~S.}\ \bibnamefont {Wiersma}},\ }\href {\doibase
  10.1364/OE.23.0A1472} {\bibfield  {journal} {\bibinfo  {journal} {Opt.
  Express}\ }\textbf {\bibinfo {volume} {23}},\ \bibinfo {pages} {A1472}
  (\bibinfo {year} {2015})}\BibitemShut {NoStop}%
\bibitem [{\citenamefont {Pratesi}\ \emph {et~al.}(2013)\citenamefont
  {Pratesi}, \citenamefont {Burresi}, \citenamefont {Riboli}, \citenamefont
  {Vynck},\ and\ \citenamefont {Wiersma}}]{Pratesi:13}%
  \BibitemOpen
  \bibfield  {author} {\bibinfo {author} {\bibfnamefont {F.}~\bibnamefont
  {Pratesi}}, \bibinfo {author} {\bibfnamefont {M.}~\bibnamefont {Burresi}},
  \bibinfo {author} {\bibfnamefont {F.}~\bibnamefont {Riboli}}, \bibinfo
  {author} {\bibfnamefont {K.}~\bibnamefont {Vynck}}, \ and\ \bibinfo {author}
  {\bibfnamefont {D.~S.}\ \bibnamefont {Wiersma}},\ }\href {\doibase
  10.1364/OE.21.00A460} {\bibfield  {journal} {\bibinfo  {journal} {Opt.
  Express}\ }\textbf {\bibinfo {volume} {21}},\ \bibinfo {pages} {A460}
  (\bibinfo {year} {2013})}\BibitemShut {NoStop}%
\bibitem [{\citenamefont {Bigourdan}\ \emph {et~al.}(2019)\citenamefont
  {Bigourdan}, \citenamefont {Pierrat},\ and\ \citenamefont
  {Carminati}}]{Bigourdan:19}%
  \BibitemOpen
  \bibfield  {author} {\bibinfo {author} {\bibfnamefont {F.}~\bibnamefont
  {Bigourdan}}, \bibinfo {author} {\bibfnamefont {R.}~\bibnamefont {Pierrat}},
  \ and\ \bibinfo {author} {\bibfnamefont {R.}~\bibnamefont {Carminati}},\
  }\href {\doibase 10.1364/OE.27.008666} {\bibfield  {journal} {\bibinfo
  {journal} {Opt. Express}\ }\textbf {\bibinfo {volume} {27}},\ \bibinfo
  {pages} {8666} (\bibinfo {year} {2019})}\BibitemShut {NoStop}%
\bibitem [{\citenamefont {Tommasi}\ \emph {et~al.}(2020)\citenamefont
  {Tommasi}, \citenamefont {Fini}, \citenamefont {Martelli},\ and\
  \citenamefont {Cavalieri}}]{TOMMASI2020124786}%
  \BibitemOpen
  \bibfield  {author} {\bibinfo {author} {\bibfnamefont {F.}~\bibnamefont
  {Tommasi}}, \bibinfo {author} {\bibfnamefont {L.}~\bibnamefont {Fini}},
  \bibinfo {author} {\bibfnamefont {F.}~\bibnamefont {Martelli}}, \ and\
  \bibinfo {author} {\bibfnamefont {S.}~\bibnamefont {Cavalieri}},\ }\href
  {\doibase https://doi.org/10.1016/j.optcom.2019.124786} {\bibfield  {journal}
  {\bibinfo  {journal} {Opt. Commun.}\ }\textbf {\bibinfo {volume} {458}},\
  \bibinfo {pages} {124786} (\bibinfo {year} {2020})}\BibitemShut {NoStop}%
\bibitem [{\citenamefont {Blanco}\ and\ \citenamefont
  {Fournier}(2003)}]{blanco}%
  \BibitemOpen
  \bibfield  {author} {\bibinfo {author} {\bibfnamefont {S.}~\bibnamefont
  {Blanco}}\ and\ \bibinfo {author} {\bibfnamefont {R.}~\bibnamefont
  {Fournier}},\ }\href@noop {} {\bibfield  {journal} {\bibinfo  {journal}
  {Europhys Lett}\ }\textbf {\bibinfo {volume} {61}},\ \bibinfo {pages} {168}
  (\bibinfo {year} {2003})}\BibitemShut {NoStop}%
\bibitem [{\citenamefont {Bardsley}\ and\ \citenamefont {Dubi}(1981)}]{dubi}%
  \BibitemOpen
  \bibfield  {author} {\bibinfo {author} {\bibfnamefont {J.}~\bibnamefont
  {Bardsley}}\ and\ \bibinfo {author} {\bibfnamefont {A.}~\bibnamefont
  {Dubi}},\ }\href@noop {} {\bibfield  {journal} {\bibinfo  {journal} {SIAM J.
  Appl. Math.}\ }\textbf {\bibinfo {volume} {40}},\ \bibinfo {pages} {71}
  (\bibinfo {year} {1981})}\BibitemShut {NoStop}%
\bibitem [{\citenamefont {Pierrat}\ \emph {et~al.}(2014)\citenamefont
  {Pierrat}, \citenamefont {Ambichl}, \citenamefont {Gigan}, \citenamefont
  {Haber}, \citenamefont {Carminati},\ and\ \citenamefont
  {Rotter}}]{Pierrat17765}%
  \BibitemOpen
  \bibfield  {author} {\bibinfo {author} {\bibfnamefont {R.}~\bibnamefont
  {Pierrat}}, \bibinfo {author} {\bibfnamefont {P.}~\bibnamefont {Ambichl}},
  \bibinfo {author} {\bibfnamefont {S.}~\bibnamefont {Gigan}}, \bibinfo
  {author} {\bibfnamefont {A.}~\bibnamefont {Haber}}, \bibinfo {author}
  {\bibfnamefont {R.}~\bibnamefont {Carminati}}, \ and\ \bibinfo {author}
  {\bibfnamefont {S.}~\bibnamefont {Rotter}},\ }\href@noop {} {\bibfield
  {journal} {\bibinfo  {journal} {PNAS}\ }\textbf {\bibinfo {volume} {111}},\
  \bibinfo {pages} {17765} (\bibinfo {year} {2014})}\BibitemShut {NoStop}%
\bibitem [{\citenamefont {Dirac}(1943)}]{Dirac}%
  \BibitemOpen
  \bibfield  {author} {\bibinfo {author} {\bibfnamefont {P.~A.~M.}\
  \bibnamefont {Dirac}},\ }\href@noop {} {\bibfield  {journal} {\bibinfo
  {journal} {British Report}\ }\textbf {\bibinfo {volume} {MS-D-5}},\ \bibinfo
  {pages} {Part I} (\bibinfo {year} {1943})}\BibitemShut {NoStop}%
\bibitem [{\citenamefont {Savo}\ \emph {et~al.}(2017)\citenamefont {Savo},
  \citenamefont {Pierrat}, \citenamefont {Najar}, \citenamefont {Carminati},
  \citenamefont {Rotter},\ and\ \citenamefont {Gigan}}]{Savo765}%
  \BibitemOpen
  \bibfield  {author} {\bibinfo {author} {\bibfnamefont {R.}~\bibnamefont
  {Savo}}, \bibinfo {author} {\bibfnamefont {R.}~\bibnamefont {Pierrat}},
  \bibinfo {author} {\bibfnamefont {U.}~\bibnamefont {Najar}}, \bibinfo
  {author} {\bibfnamefont {R.}~\bibnamefont {Carminati}}, \bibinfo {author}
  {\bibfnamefont {S.}~\bibnamefont {Rotter}}, \ and\ \bibinfo {author}
  {\bibfnamefont {S.}~\bibnamefont {Gigan}},\ }\href@noop {} {\bibfield
  {journal} {\bibinfo  {journal} {Science}\ }\textbf {\bibinfo {volume}
  {358}},\ \bibinfo {pages} {765} (\bibinfo {year} {2017})}\BibitemShut
  {NoStop}%
\bibitem [{\citenamefont {Frangipane}\ \emph {et~al.}(2019)\citenamefont
  {Frangipane}, \citenamefont {Vizsnyiczai}, \citenamefont {Maggi},
  \citenamefont {Savo}, \citenamefont {Sciortino}, \citenamefont {Gigan},\ and\
  \citenamefont {Di~Leonardo}}]{batteri}%
  \BibitemOpen
  \bibfield  {author} {\bibinfo {author} {\bibfnamefont {G.}~\bibnamefont
  {Frangipane}}, \bibinfo {author} {\bibfnamefont {G.}~\bibnamefont
  {Vizsnyiczai}}, \bibinfo {author} {\bibfnamefont {C.}~\bibnamefont {Maggi}},
  \bibinfo {author} {\bibfnamefont {R.}~\bibnamefont {Savo}}, \bibinfo {author}
  {\bibfnamefont {A.}~\bibnamefont {Sciortino}}, \bibinfo {author}
  {\bibfnamefont {S.}~\bibnamefont {Gigan}}, \ and\ \bibinfo {author}
  {\bibfnamefont {R.}~\bibnamefont {Di~Leonardo}},\ }\href@noop {} {\bibfield
  {journal} {\bibinfo  {journal} {Nat. Commun.}\ }\textbf {\bibinfo {volume}
  {10}},\ \bibinfo {pages} {2442} (\bibinfo {year} {2019})}\BibitemShut
  {NoStop}%
\bibitem [{\citenamefont {Mazzolo}(2004)}]{Mazzolo_2004}%
  \BibitemOpen
  \bibfield  {author} {\bibinfo {author} {\bibfnamefont {A.}~\bibnamefont
  {Mazzolo}},\ }\href {\doibase 10.1209/epl/i2004-10216-4} {\bibfield
  {journal} {\bibinfo  {journal} {EPL}\ }\textbf {\bibinfo {volume} {68}},\
  \bibinfo {pages} {350} (\bibinfo {year} {2004})}\BibitemShut {NoStop}%
\bibitem [{\citenamefont {Zoia}\ \emph {et~al.}(2019)\citenamefont {Zoia},
  \citenamefont {Larmier},\ and\ \citenamefont {Mancusi}}]{Zoia_2019}%
  \BibitemOpen
  \bibfield  {author} {\bibinfo {author} {\bibfnamefont {A.}~\bibnamefont
  {Zoia}}, \bibinfo {author} {\bibfnamefont {C.}~\bibnamefont {Larmier}}, \
  and\ \bibinfo {author} {\bibfnamefont {D.}~\bibnamefont {Mancusi}},\ }\href
  {\doibase 10.1209/0295-5075/127/20006} {\bibfield  {journal} {\bibinfo
  {journal} {{EPL}}\ }\textbf {\bibinfo {volume} {127}},\ \bibinfo {pages}
  {20006} (\bibinfo {year} {2019})}\BibitemShut {NoStop}%
\bibitem [{\citenamefont {Zoia}\ \emph {et~al.}(2012)\citenamefont {Zoia},
  \citenamefont {Dumonteil},\ and\ \citenamefont {Mazzolo}}]{Zoia_2012}%
  \BibitemOpen
  \bibfield  {author} {\bibinfo {author} {\bibfnamefont {A.}~\bibnamefont
  {Zoia}}, \bibinfo {author} {\bibfnamefont {E.}~\bibnamefont {Dumonteil}}, \
  and\ \bibinfo {author} {\bibfnamefont {A.}~\bibnamefont {Mazzolo}},\ }\href
  {\doibase 10.1209/0295-5075/100/40002} {\bibfield  {journal} {\bibinfo
  {journal} {{EPL}}\ }\textbf {\bibinfo {volume} {100}},\ \bibinfo {pages}
  {40002} (\bibinfo {year} {2012})}\BibitemShut {NoStop}%
\bibitem [{\citenamefont {de~Mulatier}\ \emph {et~al.}(2014)\citenamefont
  {de~Mulatier}, \citenamefont {Mazzolo},\ and\ \citenamefont
  {Zoia}}]{de_Mulatier_2014}%
  \BibitemOpen
  \bibfield  {author} {\bibinfo {author} {\bibfnamefont {C.}~\bibnamefont
  {de~Mulatier}}, \bibinfo {author} {\bibfnamefont {A.}~\bibnamefont
  {Mazzolo}}, \ and\ \bibinfo {author} {\bibfnamefont {A.}~\bibnamefont
  {Zoia}},\ }\href {\doibase 10.1209/0295-5075/107/30001} {\bibfield  {journal}
  {\bibinfo  {journal} {{EPL}}\ }\textbf {\bibinfo {volume} {107}},\ \bibinfo
  {pages} {30001} (\bibinfo {year} {2014})}\BibitemShut {NoStop}%
\bibitem [{\citenamefont {Jenkins}\ and\ \citenamefont {White}(1976)}]{jw}%
  \BibitemOpen
  \bibfield  {author} {\bibinfo {author} {\bibfnamefont {F.}~\bibnamefont
  {Jenkins}}\ and\ \bibinfo {author} {\bibfnamefont {H.}~\bibnamefont
  {White}},\ }\href@noop {} {\emph {\bibinfo {title} {Foundamentals of
  Optics}}},\ \bibinfo {edition} {4th}\ ed.\ (\bibinfo  {publisher}
  {McGraw-Hill},\ \bibinfo {year} {1976})\BibitemShut {NoStop}%
\bibitem [{\citenamefont {Zaccanti}(1991)}]{Zaccanti:91}%
  \BibitemOpen
  \bibfield  {author} {\bibinfo {author} {\bibfnamefont {G.}~\bibnamefont
  {Zaccanti}},\ }\href@noop {} {\bibfield  {journal} {\bibinfo  {journal}
  {Appl. Opt.}\ }\textbf {\bibinfo {volume} {30}},\ \bibinfo {pages} {2031}
  (\bibinfo {year} {1991})}\BibitemShut {NoStop}%
\bibitem [{\citenamefont {Contini}\ \emph {et~al.}(1997)\citenamefont
  {Contini}, \citenamefont {Martelli},\ and\ \citenamefont
  {Zaccanti}}]{Contini:97}%
  \BibitemOpen
  \bibfield  {author} {\bibinfo {author} {\bibfnamefont {D.}~\bibnamefont
  {Contini}}, \bibinfo {author} {\bibfnamefont {F.}~\bibnamefont {Martelli}}, \
  and\ \bibinfo {author} {\bibfnamefont {G.}~\bibnamefont {Zaccanti}},\ }\href
  {\doibase 10.1364/AO.36.004587} {\bibfield  {journal} {\bibinfo  {journal}
  {Appl. Opt.}\ }\textbf {\bibinfo {volume} {36}},\ \bibinfo {pages} {4587}
  (\bibinfo {year} {1997})}\BibitemShut {NoStop}%
\bibitem [{\citenamefont {Sassaroli}\ \emph {et~al.}(1998)\citenamefont
  {Sassaroli}, \citenamefont {Blumetti}, \citenamefont {Martelli},
  \citenamefont {Alianelli}, \citenamefont {Contini}, \citenamefont
  {Ismaelli},\ and\ \citenamefont {Zaccanti}}]{Sassaroli:98}%
  \BibitemOpen
  \bibfield  {author} {\bibinfo {author} {\bibfnamefont {A.}~\bibnamefont
  {Sassaroli}}, \bibinfo {author} {\bibfnamefont {C.}~\bibnamefont {Blumetti}},
  \bibinfo {author} {\bibfnamefont {F.}~\bibnamefont {Martelli}}, \bibinfo
  {author} {\bibfnamefont {L.}~\bibnamefont {Alianelli}}, \bibinfo {author}
  {\bibfnamefont {D.}~\bibnamefont {Contini}}, \bibinfo {author} {\bibfnamefont
  {A.}~\bibnamefont {Ismaelli}}, \ and\ \bibinfo {author} {\bibfnamefont
  {G.}~\bibnamefont {Zaccanti}},\ }\href@noop {} {\bibfield  {journal}
  {\bibinfo  {journal} {Appl. Opt.}\ }\textbf {\bibinfo {volume} {37}},\
  \bibinfo {pages} {7392} (\bibinfo {year} {1998})}\BibitemShut {NoStop}%
\bibitem [{\citenamefont {Zaccanti}\ \emph {et~al.}(1994)\citenamefont
  {Zaccanti}, \citenamefont {Battistelli}, \citenamefont {Bruscaglioni},\ and\
  \citenamefont {Wei}}]{Zaccantietal94}%
  \BibitemOpen
  \bibfield  {author} {\bibinfo {author} {\bibfnamefont {G.}~\bibnamefont
  {Zaccanti}}, \bibinfo {author} {\bibfnamefont {E.}~\bibnamefont
  {Battistelli}}, \bibinfo {author} {\bibfnamefont {P.}~\bibnamefont
  {Bruscaglioni}}, \ and\ \bibinfo {author} {\bibfnamefont {Q.~N.}\
  \bibnamefont {Wei}},\ }\href {\doibase 10.1088/0963-9659/3/5/019} {\bibfield
  {journal} {\bibinfo  {journal} {Pure Appl. Opt.}\ }\textbf {\bibinfo {volume}
  {3}},\ \bibinfo {pages} {897} (\bibinfo {year} {1994})}\BibitemShut {NoStop}%
\bibitem [{\citenamefont {Henyey}\ and\ \citenamefont
  {Greenstein}(1941)}]{hg2}%
  \BibitemOpen
  \bibfield  {author} {\bibinfo {author} {\bibfnamefont {L.~G.}\ \bibnamefont
  {Henyey}}\ and\ \bibinfo {author} {\bibfnamefont {J.~L.}\ \bibnamefont
  {Greenstein}},\ }\href@noop {} {\bibfield  {journal} {\bibinfo  {journal}
  {Astrophys. J.}\ }\textbf {\bibinfo {volume} {93}},\ \bibinfo {pages} {70}
  (\bibinfo {year} {1941})}\BibitemShut {NoStop}%
\bibitem [{\citenamefont {Martelli}\ \emph {et~al.}(2009)\citenamefont
  {Martelli}, \citenamefont {{Del Bianco}}, \citenamefont {Ismaelli},\ and\
  \citenamefont {Zaccanti}}]{librofabrizio}%
  \BibitemOpen
  \bibfield  {author} {\bibinfo {author} {\bibfnamefont {F.}~\bibnamefont
  {Martelli}}, \bibinfo {author} {\bibfnamefont {S.}~\bibnamefont {{Del
  Bianco}}}, \bibinfo {author} {\bibfnamefont {A.}~\bibnamefont {Ismaelli}}, \
  and\ \bibinfo {author} {\bibfnamefont {G.}~\bibnamefont {Zaccanti}},\
  }\href@noop {} {\emph {\bibinfo {title} {Light Propagation through Biological
  Tissue and Other Diffusive Media: Theory, Solutions, and Software}}}\
  (\bibinfo  {publisher} {SPIE Press/Bellingham, Whashington (USA)},\ \bibinfo
  {year} {2009})\BibitemShut {NoStop}
\end{thebibliography}
\end{document}